\documentclass[nofootinbib,aps,prl,twocolumn,reprint,amsmath,amssymb,superscriptaddress]{revtex4-2}

\usepackage{natbib}
\usepackage{siunitx}
\usepackage{placeins}
\usepackage{hyperref}
\usepackage[]{graphicx}
\usepackage[]{cleveref}
\begin{document}

\title[Supercurrent modulation in InSb nanoflag-based Josephson junctions by scanning gate microscopy]{Supercurrent modulation in InSb nanoflag-based Josephson junctions by scanning gate microscopy}

\author{Antonio Lombardi}
\affiliation{NEST, Istituto Nanoscienze-CNR and Scuola Normale Superiore, Piazza San Silvestro 12, 56127 Pisa, Italy}

\author{Gaurav Shukla}
\affiliation{NEST, Istituto Nanoscienze-CNR and Scuola Normale Superiore, Piazza San Silvestro 12, 56127 Pisa, Italy}

\author{Giada Bucci}
\affiliation{NEST, Istituto Nanoscienze-CNR and Scuola Normale Superiore, Piazza San Silvestro 12, 56127 Pisa, Italy}

\author{Sedighe Salimian}
\affiliation{NEST, Istituto Nanoscienze-CNR and Scuola Normale Superiore, Piazza San Silvestro 12, 56127 Pisa, Italy}

\author{Valentina Zannier}
\affiliation{NEST, Istituto Nanoscienze-CNR and Scuola Normale Superiore, Piazza San Silvestro 12, 56127 Pisa, Italy}

\author{Simone Traverso}
\affiliation{Dipartimento di Fisica, Università di Genova, Via Dodecaneso 33, 16146 Genova, Italy}
\affiliation{CNR-SPIN, Via Dodecaneso 33, 16146 Genova, Italy}

\author{Samuele Fracassi}
\affiliation{Dipartimento di Fisica, Università di Genova, Via Dodecaneso 33, 16146 Genova, Italy}
\affiliation{CNR-SPIN, Via Dodecaneso 33, 16146 Genova, Italy}

\author{Niccolo Traverso Ziani}
\affiliation{Dipartimento di Fisica, Università di Genova, Via Dodecaneso 33, 16146 Genova, Italy}
\affiliation{CNR-SPIN, Via Dodecaneso 33, 16146 Genova, Italy}

\author{Maura Sassetti}
\affiliation{Dipartimento di Fisica, Università di Genova, Via Dodecaneso 33, 16146 Genova, Italy}
\affiliation{CNR-SPIN, Via Dodecaneso 33, 16146 Genova, Italy}

\author{Matteo Carrega}
\email{matteo.carrega@spin.cnr.it}
\affiliation{CNR-SPIN, Via Dodecaneso 33, 16146 Genova, Italy}

\author{Fabio Beltram}
\affiliation{NEST, Istituto Nanoscienze-CNR and Scuola Normale Superiore, Piazza San Silvestro 12, 56127 Pisa, Italy}

\author{Lucia Sorba}
\affiliation{NEST, Istituto Nanoscienze-CNR and Scuola Normale Superiore, Piazza San Silvestro 12, 56127 Pisa, Italy}

\author{Stefan Heun}
\email{stefan.heun@nano.cnr.it}
\affiliation{NEST, Istituto Nanoscienze-CNR and Scuola Normale Superiore, Piazza San Silvestro 12, 56127 Pisa, Italy}


\begin{abstract}
{InSb nanoflags represent an interesting platform for quantum transport and have recently been  exploited in the study of hybrid planar Josephson junctions. Due to the uncovered semiconductor surface, they are also good candidates for surface probe techniques. Here, we report the first Scanning Gate Microscopy (SGM) experiments on Nb-contacted InSb nanoflag-based Josephson junctions. In the normal state, sizable conductance modulation via the charged tip of the SGM is recorded. 
In the superconducting state, we report the first application of Scanning Gate Microscopy to superconducting weak links, demonstrating the possibility of manipulating the supercurrent flow across a semiconductor-superconductor heterostructure at a local level. The experimental findings are consistent with theoretical predictions and establish a new way of investigating the behavior of superconducting weak links, towards the local imaging of supercurrent flow.}
\end{abstract}
%\keywords{Scanning gate microscopy, Josephson junctions, Semiconducting nanoflags}

\maketitle

%\section{Introduction}\label{sec1}
{\it Introduction.}--
In the last decades, a strong interest revolved around hybrid semiconductor/superconductor (SC) devices where the proximity effect can lead to a wide variety of phenomena. Among all, the possible emergence of topological states, such as Majorana modes, triggered huge efforts starting from the first proposal based on hybrid SC-semiconducting nanowires with strong spin-orbit coupling \cite{Kitaev,Stern,rohtua,prada}. More recently, planar Josephson junctions, where a 2D semiconductor is placed in between two SCs forming a super-normal-super (SNS) junction, emerged as an alternative platform to achieve non-trivial topological phases \cite{Ren,Fornieri,Lesser,Pientka}.

Interestingly, in planar junctions, superconductors can be contacted laterally, retaining the semiconducting surface of the device uncovered, allowing for the investigation by spatially resolved techniques. Among all surface probe techniques, in the last twenty years Scanning Gate Microscopy (SGM) has proven itself to be successful in the study of semiconductor-based devices \cite{Eriksson,Topinka,Topinka2,Jura,Jura2,LeRoy,Percebois,Gold, Gold2,Aidala, Paradiso,PARADISO3, paradiso2, Tomimatsu,Braem2,Bours,Aharon,Marguerite,Dou,Sakanashi,Bachtold,Freitag,Bozovic,Denisov,Gilde,Huefner,Huefner2,Martins,Fan,Bhandari,Hegedues2025,Maji2024,Maji2025}. This technique combines surface and transport studies: an atomically sharp tip is used to perform local gating on a semiconductor. %Local gating means that one can locally change the carrier density and the electrostatic potential felt by the electrons below the tip. Consequently, by scanning the sample surface and measuring the electron transport, one can link quantities like conductance (or critical current) to the tip position. In this way, a movable local scattering centre for charge carriers is obtained, whose spatial dimensions vary depending on tip diameter, tip-to-sample distance, and potential difference. 
By current or voltage biasing the sample and recording its conductance response as a function of tip position, information can be obtained on the current density distribution in the sample as well as the presence, nature, and location of defects \cite{Paradiso,paradiso2,PARADISO3,Iagallo,brun_kondo}. Electron transport in quantum wells \cite{Eriksson,Topinka2,Topinka,Jura,Jura2,Gold,Aidala,Paradiso,paradiso2,PARADISO3,Tomimatsu,Braem2}, two-dimensional materials like graphene \cite{Bours,Aharon,Gold2,Marguerite,Dou} and MoTe$_2$ \cite{Sakanashi}, carbon nanotubes \cite{Bachtold,Freitag,Bozovic}, quantum dots \cite{Denisov,Gilde,Huefner}, quantum rings \cite{Martins}, and nanowires \cite{Fan,ania} has been investigated by SGM.

To date, only three SGM experiments have been performed on superconducting devices so far. In 2009, Huefner et al.~\cite{Huefner2} investigated a superconducting single electron transistor based on an Al island, in a regime where no Josephson effect was present across the tunnel junctions connecting the dot to the electrodes. In 2020, Bhandari et al.~\cite{Bhandari} imaged Andreev reflection in graphene in an architecture with a single superconducting electrode, by exploiting cyclotron motion to spatially separate the particles impinging on the electrode from those reflected by it. In 2024, Hegedüs et al.~\cite{Hegedues2025} studied the effect of two-level defects as noise sources in superconducting qubits. To our knowledge there is so far no record of SGM experiments performed on superconducting weak links (SNS), such as hybrid SC-semiconductor Josephson junctions. Such hybrid structures have been intensively investigated by transport techniques, but it is still an open challenge to map and manipulate individual features in their supercurrent density distribution, e.g., Josephson vortices. Furthermore, superconducting junctions based on materials with large spin-orbit coupling (e.g., InSb) promise to host exotic topological phases, whose local manipulation would be a key step in the route towards topological quantum computation \cite{Stern}. Therefore, assessing to local supercurrent density and visualizing electronic flow in hybrid devices by SGM can give relevant information also in the context of topological platforms.

We note that recently an alternative way of investigating supercurrent flow by scanning probe microscopy has been demonstrated: Chen et al.~mapped the magnetic field distribution generated by the supercurrent in a Josephson junction via a scanning probe magnetometer based on a single nitrogen vacancy in diamond \cite{Chen2}. Despite its undeniable potential, which includes the possibility of mapping the supercurrent distribution in conditions different from criticality, this technique requires samples with high critical current densities to create a sufficiently strong magnetic field to be detected. 

InSb nanoflags have recently sparked interest as a quasi-two-dimensional alternative to InSb nanowire-based devices \cite{Isha, Prosko}. Several groups have investigated high mobility InSb nanoflag-based devices, demonstrating gate tunable proximity-induced superconductivity \cite{Yan,DEV,Sedi} as well as electrostatic control over spin-orbit interaction \cite{Kang, Chen}, superconducting diode effect \cite{Bianca}, and half-integer Shapiro steps \cite{Iorio}. In these devices, the normal semiconducting region between the superconducting electrodes remains exposed, which opens the possibility for the application of the SGM technique.

In this paper we report the first SGM measurements of InSb nanoflag-based Josephson junctions with Nb contacts, both in the normal and in the superconducting regime. These measurements represent both the first SGM experiment on InSb nanoflag and, at the same time, the first SGM report on SNS devices, demonstrating local critical current modulation via the charged tip of the SGM.

%\section{Results}\label{sec2}
{\it Experimental setup and scanning gate microscopy.}--
Measurements have been acquired in a Janis bath cryostat equipped with a $^3$He refrigeration stage with a base temperature of $\SI{300}{mK}$ and a custom-made Attocube Scanning Gate Microscope. Both the electrical transport lines and the connections necessary for the operation of the microscope have been filtered to preserve the superconducting behavior of the devices under study \cite{Guiducci-m,Guiducci2019}. 

\begin{figure}[tbp]
	\centering
	\includegraphics[width=\linewidth]{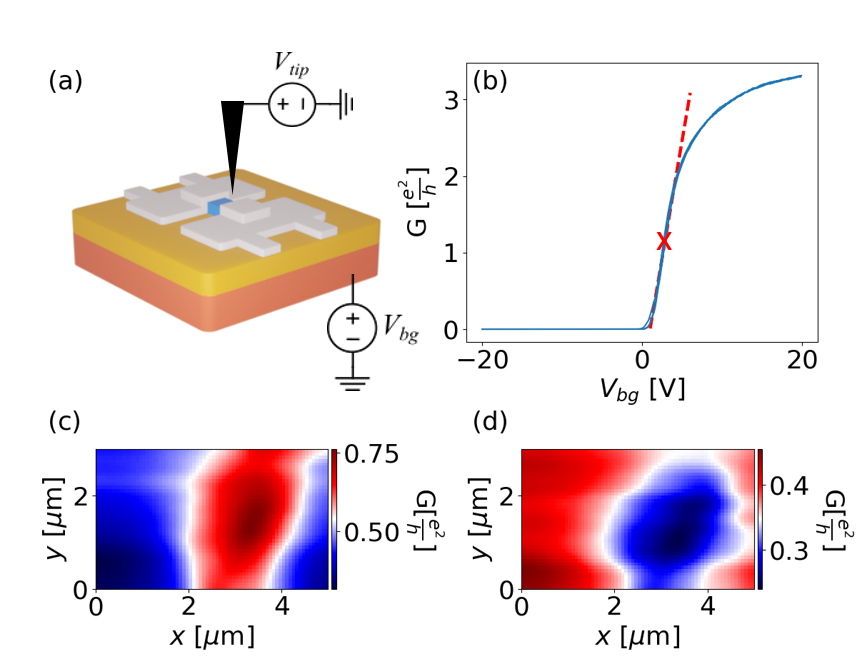}
	\caption{\label{fig:1} (a) Graphical representation of a Scanning Gate Microscopy experiment. The light blue section corresponds to the exposed semiconducting region of the nanoflag. (b) Conductance modulation of device SC6 versus back gate voltage. The back gate working point $V_{bg} = \SI{3}{V}$ is indicated by a red x. The numerical least squares fit of the rising slope in proximity to the working point with a linear model is indicated by the red dashed line. (c) Conductance map as a function of tip position acquired in dual pass mode	with a $\SI{60}{nm}$ tip offset with respect to the topography scan. Working point:	$V_{tip} = \SI{10}{V}$, $V_{bg} = \SI{3}{V}$. (d) Conductance map acquired in the same conditions as in (c), but with $V_{tip} = -\SI{10}{V}$.}
\end{figure}

A typical device design is reported in Fig.~\ref{fig:1}(a): it consists in a $\sim \mu$m-sized InSb nanoflag deposited on a highly conductive p-type Si(100) substrate, which serves as a global back gate.  A $\SI{285}{nm}$ thick SiO$_2$ layer covers the Si substrate as dielectric. The nanoflag is contacted through two Nb electrodes \cite{Isha,Sedi,Bianca,Iorio}, separated by a gap about $\SI{200}{nm}$ long, i.e., the semiconducting region of the Josephson junction. Electrostatic control of the global behavior of the junction can be achieved by polarizing the back gate electrode with a voltage difference $V_{bg}$ with respect to the grounded device. In the experiments, we have investigated the action of an additional gate electrode, the SGM tip, once it is placed in proximity to the junction and polarized with a voltage difference $V_{tip}$ with respect to it.
Details on the device fabrication are provided in section S1 of the Supplemental Material (SM).

First measurements were performed at $\SI{77}{K}$. At this temperature, the devices behave as conventional n-type field-effect transistors, as displayed by the conductance modulation curve shown in Fig.~\ref{fig:1}(b). When the charged tip is placed close to the exposed semiconducting region of the junction, it modifies the local carrier concentration. This effect can be recorded by measuring the conductance across the device as a function of the tip position. SGM measurements have been performed in dual pass mode (see S2 of the SM) by acquiring the in-phase component of the voltage difference across the junction via lock-in technique.

In order to enhance the effect of the tip, as working point a back gate potential of $V_{bg} = \SI{3}{V}$ has been set, where the change of sample conductance with back gate voltage is maximized, as shown in Fig.~\ref{fig:1}(b). In Fig.~\ref{fig:1}(c) we report a conductance modulation map induced by a tip voltage of $V_{tip} = \SI{10}{V}$. The location of the exposed region of the nanoflag is shown as a red area in the conductance map: once the positively charged tip gets sufficiently close to the device, it induces charge carrier accumulation in the InSb nanoflag, thereby enhancing its conductance. To demonstrate that this effect is caused by tip-induced charge carrier density modulation, SGM maps have been acquired with both positive and negative voltages applied to the tip (see Fig.~\ref{fig:1}(c-d)). As expected, while the positive polarization of the tip induces conductance enhancement, the negative polarization induces conductance reduction (blue area in Fig.~\ref{fig:1}(d)) once the tip is placed close to the exposed semiconducting region. We note that in both Fig.~\ref{fig:1}(c) and Fig.~\ref{fig:1}(d) the junction conductance recorded in the SGM scans is smaller than that corresponding to the working point in Fig.~\ref{fig:1}(b). We attribute this effect to a shift in the position of the conductance modulation curve caused by the presence of the tip close to the device. In fact, the change in both tip-to-sample distance and $V_{tip}$ involved in dual pass operation can trigger charge-redistribution effects. However, the presence of this effect does not affect the analysis nor the interpretation of the data, since it affects all measurements in the same way.

The effect of tip bias $V_{tip}$ has been studied by performing dual pass scans with a lift-off height of $\SI{60}{nm}$ with respect to the reference topography, see section S3 of the SM. From these measurements, we extract a tip gating efficiency $\alpha_{tip} = dG/dV_{tip} = 0.047\pm \SI{0.008}{\frac{e^2}{hV}}$ at $V_{bg}=\SI{3}{V}$, defined as the change in device conductance with respect to the variation in tip-sample voltage difference. Furthermore, we have extracted the back gate efficiency $\alpha_{bg}= dG/dV_{bg}$ with the tip placed far away from the semiconducting region, i.e., the slope of the linear region of the opening curve of the device, as $\alpha_{bg} = 0.62\pm \SI{0.05}{\frac{e^2}{hV}}$, see the red dashed line in Fig.~\ref{fig:1}(b). This results in a ratio of gating efficiencies between the tip and the back gate of $\alpha_{tip} / \alpha_{bg} = 0.08 \pm 0.02$. This indicates that, when placed at $\SI{60}{nm}$ distance from the sample, the effect of $\SI{10}{V}$ on the tip is equivalent to a variation of the back gate potential of about $\SI{0.8}{V}$.

One of the main difficulties found in performing SGM measurements on these samples was related to the strong susceptibility of the system to electric perturbations. %Due to material pickup by the tip, the response time of the feedback loop got worse over time, leading to the charged tip occasionally getting into contact with one of the Nb leads. This electrical spike can change the electrostatics of the nanoflag, shifting its back gate threshold voltage and therefore the position of the chosen working point on the conductance back gate modulation curve. To avoid such accidents, 
Therefore, for the data show below we decided that $\SI{100}{nm}$ is the optimal safety distance between tip and sample.

\begin{figure}[tbp]
	\centering
	\includegraphics[width=\linewidth]{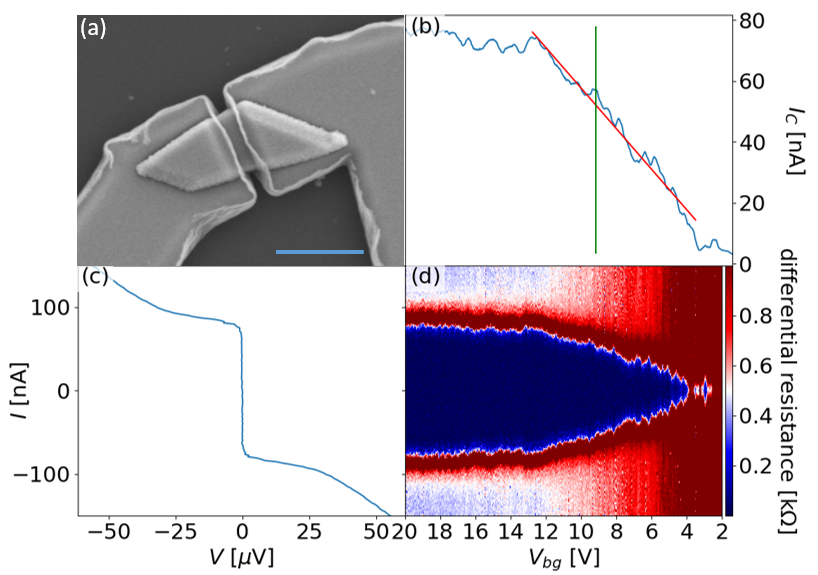}
	\caption{\label{fig:3} (a) Scanning Electron Microscopy (SEM) image of device SGM1 D8D2. The blue scalebar corresponds to $\SI{1}{\mu m}$ (b) Critical current as a function of back gate voltage. The green line indicates the position of the working point $V_{bg} = \SI{9.5}{V}$. The red line shows to the best linear fit of the critical current modulation by the back gate in the neighborhood of the working point. (c) Current-voltage curve of device SGM1 D8D2 at $V_{bg} = \SI{20}{V}$. (d) Differential resistance map as a function of back gate voltage and current bias.}
\end{figure}

{\it Supercurrent Flow Manipulation.}-- 
Mapping supercurrent profiles with SGM on SNS devices has been theoretically proposed recently \cite{kaperek}. Here, we report the first experimental demonstration of SGM on superconducting weak links. The device is displayed in Fig.~\ref{fig:3}(a). At $\SI{300}{mK}$, we report an open channel switching current of about $\SI{80}{nA}$ (see Fig.~\ref{fig:3}(c)), a value consistent with the ones reported from previous experiments on similar structures \cite{Sedi,Bianca}. As the back gate voltage is reduced, we identified three regimes in the junction behavior (see Fig.~\ref{fig:3}(d)): (i) For $V_{bg} > \SI{14}{V}$, no back gate induced critical current modulation is present, similar to what was reported in \cite{Sedi}. (ii) For $\SI{14}{V} > V_{bg} > \SI{2.5}{V}$, the critical current experiences an overall downward trend. We have identified this as the optimal region to place the back gate set-point in order to maximize the influence of the tip on the behavior of the junction. (iii) For $V_{bg} < \SI{2.5}{V}$, the Josephson effect is completely suppressed in the weak link. Thus, the device works as a Josephson field-effect transistor (JoFET).
 
We identified the optimal working point as $V_{bg} = \SI{9.5}{V}$ and performed a least squares fit with a linear model of the critical current modulation by the back gate voltage close to the working point, as shown in Fig.~\ref{fig:3}(b). We obtain a critical current modulation efficiency $\beta_{bg} =\delta I_c/\delta V_{bg}= \SI{6.5}{\frac{nA}{V}}$.

\begin{figure*}[tbp]
	\centering
	\includegraphics[width=\linewidth]{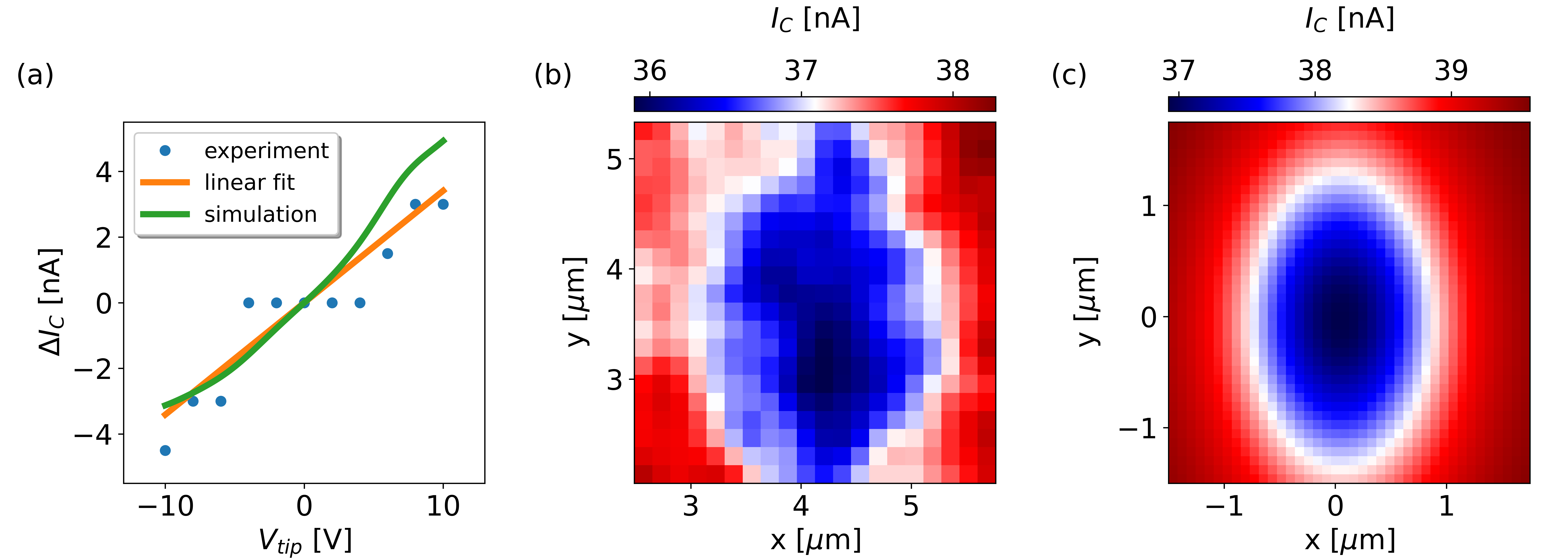}
	\caption{\label{fig:4}
	(a) Critical current as a function of tip-to-sample voltage difference $V_{tip}$ when the tip is placed close to the geometrical centre of the junction; the best linear fit to the displayed data is shown in orange, while the result of numerical simulations is reported in green. (b) SGM map measurement: $20$ pixel $\times$ $20$ pixel critical current map as a function of the position of the tip. $V_{tip} = \SI{-10}{V}$, $V_{bg} = \SI{9.5}{V}$. (c) Numerical simulation of tip-induced critical current modulation map (details on the theoretical model can be found in S4 of the Supplemental Material).}
\end{figure*}

The SGM tip has then been placed in the middle of the junction at a safety distance of $\SI{100}{nm}$ above the highest point recorded in a previously acquired AFM topography map (for the reasons discussed above). Then, keeping the tip position fixed, we have acquired $I-V$ curves in a four-probe DC current bias configuration for different values of the tip potential. The resulting variation in critical current as a function of the tip-to-sample voltage difference is reported in Fig.~\ref{fig:4}(a) as blue dots, showing a clear reduction in critical current when $V_{tip}<0$ and an enhancement when $V_{tip}>0$.

The observed modulation in critical current is of a few nA, which is in agreement with numerical simulations, see the green line in Fig.~\ref{fig:4}(a). For details on the simulations, see below and S4 in the SM. By performing a least squares fit with a linear model, we obtain a critical current modulation efficiency $\beta_{tip} = \delta I_C/\delta V_{tip}= \SI{0.35}{\frac{nA}{V}}$ (orange line in Fig.~\ref{fig:4}(a)). This indicates that, in these conditions, the effect of applying $\SI{10}{V}$ on the tip is equivalent to a variation in the back gate voltage of about $\SI{0.5}{V}$, which is roughly consistent with the value obtained at $\SI{77}{K}$. Such trends have been reproduced over multiple acquisitions, confirming that the tip exerts an influence on the superconductive behavior of the sample, opening up the possibility of performing SGM critical current maps.

Next, the tip has been scanned in an evenly spaced grid of 20 x 20 positions while maintaining a constant vertical distance of $\SI{100}{nm}$; we have kept a fixed voltage difference $V_{tip} = \SI{-10}{V}$  with respect to the sample, in order to induce depletion of charge carriers. For each grid-point we have acquired an $I-V$ curve of the device in a four-wire DC current bias setup, using a NiDAQ acquisition board for both signal generation and acquisition at its maximum sampling frequency of $f = \SI{5}{kHz}$. From this data, the critical current for each grid-point was determined. The results are reported in Fig.~\ref{fig:4}(b). The map shows a clear reduction of the supercurrent of the device, indicated by the blue area in the middle of the plot, once the tip is placed close to the location of the junction. As the tip-to-junction lateral distance increases, an increase in the critical current is observed, indicating the local nature of the tip-device interaction. We observe a $\Delta I_C = \SI{2}{nA}$ critical current signal difference between the center and the margins of the scan area: this is consistent with the measurements reported in Fig.~\ref{fig:4}(a); we can explain the difference between $\Delta I_C$ and its expected value $\Delta I_{C0} = \SI{10}{V}\cdot\beta_{tip} = \SI{3.5}{nA}$ with the presence of a residual electrostatic interaction between the device and the tip when it is placed at the margins of the scan range. In Fig.~\ref{fig:4}(c) we report the results of a numerical simulation of the tip-induced critical current modulation, which reliably reflect all the features observed in the experimental data. Details on the theoretical model and simulation parameters can be found in S4 of the SM.
We underline that consistent results were obtained on a second device, see section S5 of the SM.

%\section{Discussion}\label{sec12}
{\it Discussion and conclusions.}--
We have reported the first SGM experiments performed on both InSb nanoflag-based devices and superconducting weak links. In particular, we have reported the first SGM critical current modulation maps, demonstrating the possibility of locally manipulating the behavior of superconductive weak links through a charged SGM tip. This last achievement is a proof of principle that will set a novel path for future experiments to characterize and manipulate the superconducting behavior of hybrid systems at a local level.

In order to progress in the development of the SGM technique, it is necessary to improve its spatial resolution. Furthermore, the tip-to-sample capacitive coupling decreases with increasing sample-to-tip distance, limiting the experiment signal-to-noise ratio. Therefore, it is of pivotal importance for the future development of the technique to find ways of reducing this distance.

This can be achieved by depositing on the region of interest a few-nm thick layer of high performance dielectric (e.g.~HfO$_2$) by atomic layer deposition. This will allow to perform SGM with the polarized tip placed directly in contact with the dielectric layer while at the same time preventing the danger of electrical discharges and increasing the tip-to-sample capacitance. Once these developments come into play, it will be possible to acquire well-resolved critical current modulation maps, opening up a way to perform supercurrent flow imaging, to better understand the physics of superconducting weak links.

Finally, it is worth mentioning that our study opens up a promising way of investigating and manipulating topological phases at a local level via SGM, a necessary step in the path to engineering devices suitable to perform fault-tolerant quantum computation. In fact, the presence of the polarized tip can be used to locally deplete a topological device, inducing a local topological phase transition, and leading to the formation of Majorana zero modes.

%\section{Methods}\label{sec11}
\

\begin{acknowledgments}
Useful discussions with Michal P. Nowak and Fernando Domingues are gratefully acknowledged. The authors acknowledge the support from the project PRIN2022 2022-PH852L(PE3) TopoFlags—“Non reciprocal supercurrent and topological transition in hybrid Nb-InSb nanoflags” funded by the European community—Next Generation EU within the program “PNRR Missione 4—Componente 2—Investimento 1.1 Fondo per il Programma Nazionale di Ricerca e Progetti di Rilevante Interesse Nazionale (PRIN)” and by PNRR MUR Project No. PE0000023-NQSTI.
\end{acknowledgments}


\bibliography{bib3}% common bib file

\end{document}